\documentclass[doublecol]{epl2}
\usepackage{bm}
\usepackage{graphicx}
\usepackage{amssymb,amsmath}
\usepackage{color}
\usepackage{hyperref}

\newcommand{\p}{^\prime}
\newcommand{\T}{\tau}

\newcommand{\gf}{G_{f}(\T)}
\newcommand{\gfi}{G_{f}(i\omega_n)}
\newcommand{\gx}{G_{X}(\T)}
\newcommand{\gxi}{G_{X}(i\nu_n)}

\title{Magnetic fluctuations near the Mott transition towards a spin liquid state}
\shorttitle{Magnetic fluctuations near the Mott transition towards a spin liquid state}
\author{Serge Florens \inst{1}, Priyanka Mohan \inst{2}, C. Janani \inst{2}, T.
Gupta \inst{3,4}, R. Narayanan \inst{2}}
\institute{ 
\inst{1} Institut N\'{e}el, CNRS and UJF, B.P. 166, 25 Avenue des Martyrs, 38042 Grenoble 
Cedex 9, France\\
\inst{2} Department of Physics, Indian Institute of Technology, Chennai-600036, India\\
\inst{3} The Institute of Mathematical Sciences, C.I.T. Campus, Chennai 600 113, India\\
\inst{4} Theoretical Physics Division, Indian association for the Cultivation of
Sciences, Kolkata, India
}


\abstract{
Dynamics of magnetic moments near the Mott metal-insulator transition is
investigated by a combined slave-rotor and Dynamical Mean-Field Theory solution of 
the Hubbard model with additional fully-frustrated random Heisenberg couplings.
In the paramagnetic Mott state, the spinon decomposition allows to generate a 
Sachdev-Ye spin liquid in place of the collection of independent local moments
that typically occurs in the absence of magnetic correlations.
Cooling down into the spin-liquid phase, the onset of deviations from pure Curie behavior 
in the spin susceptibility is found to be correlated to the temperature scale at
which the Mott transition lines experience a marked bending.
We also demonstrate a weakening of the effective exchange energy upon approaching the 
Mott boundary from the Heisenberg limit, due to quantum fluctuations associated to zero 
and doubly occupied sites.
}

\begin{document}

\maketitle

\email{serge.florens@grenoble.cnrs.fr}

\section{Introduction}
The Mott metal-insulator transition, wherein electronic waves are localized 
by short-range electron-electron interactions (see \cite{Imada98} for a review), is 
one of the most complex phenomenon observed in strongly correlated electronic
systems.
Even though the appearance of a Mott gap is purely driven by the charge degrees 
of freedom, it is expected that magnetic fluctuations play a very crucial role in 
determining the true nature of this phase transition.
In the paramagnetic Mott insulator, local moments are indeed well defined 
objects after their creation at high temperature (at a scale set by the local Coulomb 
interaction) and before their ultimate antiferromagnetic ordering at the N\'eel temperature, 
offering a window in which complex behavior of the spin excitations is yet 
to be clearly understood. 
Experimentally, the simplest situation in this respect occurs when the low-temperature 
magnetic ordering is first order, as in the case of Cr-doped V$_2$O$_3$. Since the magnetic 
correlations are expected to be weak in this case, many predictions can be made from a single-site 
approach like the Dynamical Mean Field Theory (DMFT) \cite{Georges96}, where local moments 
are described as freely fluctuating from each other in the insulating state. 
As a consequence, the low-temperature metallic phase leads upon heating to an insulator, giving 
a Pomeranchuk-like effect where the entropy gain benefits to the state with magnetic degeneracy. 
Conversely, other classes of materials, such as the $\kappa$-organics, display a
continuous magnetic transition into the N\'eel state, so that the localized spins will experience 
strong collective fluctuations. The most striking experimental consequence
\cite{Lefebvre00,Limelette03,Kagawa04} lies in the progressive disappearance
of the Pomeranchuk effect upon approaching the magnetic ordering temperature,
so that the Mott transition lines in the pressure-temperature phase diagram
{\it bend} at low temperature. 
These qualitative arguments have received recent confirmation from cluster DMFT calculations 
of the phase diagram of the two-dimensional Hubbard model on a frustrated lattice 
\cite{Ohashi08,Park08,Balzer08,Liebsch09}, that take into account nearest neighbor
magnetic exchange leading to short-range singlet formation. 
However, the precise connection between the appearance of low-energy magnetic excitations and
deviations from the phase diagram of the single-site approach has remained unclear.

Another challenging aspect of correlated fermion materials, that we wish to
address here, concerns the interplay with disorder, that can lead to a plethora of
fascinating physics, such as Anderson-Mott transitions~\cite{Vlad} and highly frustrated 
spin-liquid phases~\cite{Sachdev93,Parcollet99}, for which extensions of Dynamical Mean Field 
ideas can also bring interesting light. In particular, introducing magnetic frustration
(for instance due to random magnetic couplings) can have a profound influence on the nature of
the Mott transition, by modifying the entropy of the magnetic Mott state, and
also giving rise to slow spin dynamics~\cite{RSidd} that can couple non-trivially to
electronic excitations in the metallic phase as
well~\cite{Parcollet99,Rosch,Miranda}.
The real nature of the Mott transition at zero temperature is believed infact to be 
strongly influenced by the type of magnetic correlations built by exchange
interactions, so that the standard DMFT scenario~\cite{Georges96} of a second
order Brinkman-Rice transition may not apply anymore. However, due the complexity of
the impurity models, which involve both fermionic and bosonic baths (see
Eq.~(\ref{action}) below), such studies that combine Mott physics with random exchange have been
rare~\cite{Parcollet99}. One of our aim here is to provide a first attempt at a fully 
dynamical solution of this complex physical problem.

\section{Methodology}
In order to investigate the interplay of Mott physics and magnetic fluctuations
in a controlled setting, we consider the standard Hubbard model supplemented with long-range 
fully-frustrated and random Heisenberg exchange interactions, taking frustration as a 
necessary prerequisite for a non-trivial spin liquid ground
state~\cite{Hassan13} on the insulating side of the Mott transition:
\begin{eqnarray}
\nonumber
H &=& - t \sum_{\langle i\,j\rangle,\,\sigma} d^{\dagger}_{i\,\sigma}
d_{j\,\sigma}^{\phantom{\dagger}} 
- \mu \sum_{i\,\sigma}d^{\dagger}_{i\,\sigma} d_{i\,\sigma}^{\phantom{\dagger}} \\
&& + U \sum_{i} d^{\dagger}_{i \uparrow} d_{i\uparrow}^{\phantom{\dagger}}
d^{\dagger}_{i\,\downarrow}d_{i\downarrow}^{\phantom{\dagger}} 
+ \sum_{ij} J_{ij} \vec{S_i}. \vec{S_j}.
\label{model}
\end{eqnarray}
Note that hopping parameters $t$ run over adjacent sites of a yet
arbitrary lattice, while the exchange terms $J_{ij}$ connect all the local
spin variables $\vec{S_i} = \sum_{\sigma,\sigma'} d^{\dagger}_{i\,\sigma}
\frac{\tau_{\sigma,\sigma'}}{2} d_{i\,\sigma'}^{\phantom{\dagger}}$ in a random fashion (amplitude and 
sign). Also, in Eq.~(\ref{model}), $\mu$ is the chemical potential and $U$ is the local on-site 
repulsion, which controls the degree of electronic correlations.

This non-trivial Hamiltonian greatly simplifies in the large coordination limit, 
where all the quantum dynamics can be exactly encapsulated by a single-site action:
\begin{eqnarray}
\label{action}
\mathcal{S} &=& \int^\beta_0 \!\!\! d\tau \left[ \sum_{\sigma}
d^{\dag}_{\sigma}(\partial_{\tau}-\mu)d_{\sigma}^{\phantom{\dagger}}
+
Ud^{\dag}_{\uparrow}d_{\uparrow}d^{\dag}_{\downarrow}d_{\downarrow}^{\phantom{\dagger}} \right]\\
\nonumber 
&& + \int^{\beta}_0 \!\!\! d\tau \int_0^{\beta} \!\!\!\!\! d\tau\p 
\Delta(\tau-\tau\p) \sum_{\sigma}d^{\dag}_{\sigma}(\tau)
d_{\sigma}^{\phantom{\dagger}}(\tau\p)\\
\nonumber 
&& + \int^{\beta}_0 \!\!\! d\tau \int_0^{\beta} \!\!\!\!\! d\tau\p 
Q(\tau-\tau') \vec{S}(\tau).\vec{S}(\tau').
\end{eqnarray}
Here, two different baths that depend on imaginary-time $\tau$ have been
introduced ($\beta=1/T$ is the inverse temperature): 
i) hopping processes from the local site into the lattice (and back)
result in a retarded hybridization $\Delta(\tau-\tau\p)$, which corresponds 
to the standard DMFT single site action \cite{Georges96}; ii) non-local
magnetic coupling exactly amounts to a retarded spin interaction $Q(\tau-\tau')$, 
providing friction on the local moments, hence reducing their entropy. The 
latter term is easily obtained through a Gaussian averaging over the disordered 
exchange coupling $J_{ij}$ in Eq.~(\ref{model}), followed by a dynamical decoupling 
of a four-spin interaction, as shown in \cite{Bray80,Sachdev93}.
For a Bethe lattice, both baths are determined self-consistently, through the rather 
simple set of equations:
\begin{eqnarray}
\label{selfconsistency1}
\Delta(\tau) &=& t^2 G_d(\tau)\\
Q(\tau) &=& J^2 \chi(\tau).
\label{selfconsistency2}
\end{eqnarray}
Here, $G_d(\T) \equiv -\langle d_{\sigma}^{\phantom{\dagger}}(\tau) 
d^{\dag}_{\sigma}(0)\rangle$ is the 
local $d$-electron single-particle Green's function, and
$\chi(\tau) \equiv \langle \vec{S}(\tau).\vec{S}(0)\rangle$ the local $d$-electron 
spin susceptibility. Free $d$-electrons (for $U=0$) are endowed with a semi-circular local 
density of states with half-width $D=2t$, while the degree of magnetic frustration 
is controlled by $J$, the width of the probability distribution used to perform the 
disorder average.

We now solve the still complicated quantum impurity problem
(\ref{action}), by laying recourse to a method that captures all the crucial physical
ingredients at play. A simple possibility is to enlarge the spin symmetry from SU(2)
to SU($N$) in the large $N$ limit \cite{Auerbach94}, from which the good emergent variables in the
magnetic sector are known \cite{Sachdev93,Parcollet99} to be spin-carrying zero-charge 
fermions (dubbed spinons) $f^{\dag}_{\sigma}$, with $\sigma=1\ldots N$ (we
comment below on the validity of this approach for realistic fermion species
with $N=2$).
Charge is then naturally encoded through a second auxiliary field, the phase $\phi$ dual to 
the local charge, resulting in the so-called slave-rotor representation
\cite{Florens02,Florens04} of the physical electron: $d^{\dag}_{\sigma} = f^{\dag}_{\sigma} e^{i \phi}\equiv
f^{\dag}_{\sigma} X$. Following \cite{Florens02}, we have written the phase as a 
complex bosonic field $X$ constrained as $|X(\tau)|^2=1$ through a Lagrange multiplier
$\Lambda(\tau)$, so that the action reads (particle-hole symmetry, i.e.  $\mu=-U/2$,
is assumed here):
\begin{eqnarray}
\label{rotor}
 \mathcal{S} &=& \int_0^\beta \!\!\! d\T \left[\frac{|\partial_\T X|^2}{4U} +
\Lambda(\T)(|X|^2-1)
+ \sum_{\sigma} f^\dag_\sigma \partial_{\tau} f_\sigma^{\phantom{\dagger}}\right]\\
\nonumber
 &+& \int^{\beta}_0 \!\!\! d\tau\int_0^{\beta} \!\!\!\!\! d\tau\p
\Delta(\tau-\tau\p) \sum_{\sigma}f^{\dag}_{\sigma}(\tau)
X(\T)f_{\sigma}^{\phantom{\dagger}}(\tau\p) X^\dagger(\T\p)\\
\nonumber
&+& \int_0^\beta \!\!\! d\tau \int_0^\beta \!\!\!\!\! d\T\p \frac{Q(\tau-\tau\p)}{N}
\sum_{\sigma\, \sigma\p} 
f^\dag_\sigma(\tau) f_{\sigma\p}^{\phantom{\dagger}}(\tau)
f^\dag_{\sigma\p}(\tau\p)f_\sigma^{\phantom{\dagger}}(\tau\p).
\end{eqnarray}
In the large $N$ limit of \cite{Florens02}, the Green's functions
for the auxiliary particles are readily obtained from Dyson equation in
Matsubara frequency:
\begin{eqnarray}
\label{GX}
G_X^{-1}(i\nu_n) = \frac{{\nu_n}^2}{U} + \Lambda - \Sigma_X(i\nu_n)\\
G_f^{-1}(i\omega_n) = i\omega_n - \Sigma_f (i\omega_n),
\label{Gf}
\end{eqnarray}
where the associated self-energies read in imaginary time:
\begin{eqnarray}
\label{SX}
\Sigma_X(\T)&=& N\Delta(\T)G_f(\T)\\
\label{Sf}
\Sigma_f(\T) &=& \Delta(\T)G_X(\T)+ Q(\T)G_f(\T).
\end{eqnarray}
Similarly, the parameter $\Lambda$ in Eq.~(\ref{GX}) is obtained by the
average constraint $G_X(\tau=0)=1$.
When re-expressed in terms of the emergent variables, the baths entering the self-consistency 
conditions Eqs.~(\ref{selfconsistency1}-\ref{selfconsistency2}) can be simplified as:
$\Delta(\T)=t^2 G_d(\T)=t^2G_f(\T)G_X(\T)$ and $Q(\T)=J^2\chi(\T)=J^2G_f(\T)^2$.
Together with these self-consistent equations, the above self-energies form a
non-linear set of integral equations that we solve numerically. We note that two
important limits are naturally incorporated in this scheme: $i)$ For $J=0$ (standard
Hubbard model), the spin bath $Q(\tau)$ can be dropped, and we recover the slave-rotor 
treatment \cite{Florens02} of the usual DMFT equations, so that a Mott
metal-insulator transition occurs at a critical value of $U/t$; $ii)$ More interestingly at
$U=\infty$, charge fluctuations are fully suppressed, as the gapped rotor
propagator $G_X(\tau)$ can be discarded in Eq.~(\ref{Sf}). One then recovers
the quantum spin liquid equations of Sachdev and Ye \cite{Sachdev93}, where
the exchange $J$ is responsible for changing the high temperature Curie behavior
of the spin susceptibility $\chi(\tau)\simeq 1/4$ into the slow temporal decay
$\chi(\tau)\propto 1/|\tau|$. As a consequence, the temperature dependence of the static
susceptibility $\chi(T)=\int_0^\beta \! d\tau \chi(\tau)$ turns from a Curie law 
$\chi(T)=1/(4T)$ at $T\gg J$ into a slower logarithmic divergence 
$\chi(T)\propto \log(1/T)$ at $T\ll J$. This interesting absence of (spin glass) magnetic 
ordering at finite temperature, due to the enhanced quantum fluctuations brought by the
large $N$ fermionic representation of the spin, allows us to study
the onset of magnetic fluctuations in the whole range of temperatures.
Some comments are in order here on the validity of the large $N$ limit for describing realistic
single-band Heisenberg spin-liquid models at $N=2$.
Indeed, pioneer quantum Monte Carlo simulations~\cite{Grempel} of the quantum
spin glass model (in the $U=\infty$ limit) reported a very different 
class of solutions at finite temperature, showing a renormalized finite Curie constant, hence 
a still $1/T$ diverging spin susceptibility, instead of the $\log(1/T)$ result obtained by 
Sachdev and Ye at large-$N$~\cite{Sachdev93}. Further studies~\cite{Arrachea,GPS,VBS} 
demonstrated however that these paramagnetic solutions were actually metastable states 
below the freezing temperature. The correct low temperature physics at $N=2$ is infact 
the one of a nearly quantum critical state (with tiny Edwards-Anderson order
parameter $q_{\mathrm{EA}}$=0.06), where low-energy excitations are perfectly described by 
the large-$N$ limit. 
In addition, we also note that the increase of quantum fluctuations occuring in our
finite $U$ Hubbard model has the tendency to enhance spin-liquid behavior, 
since we are far from the Heisenberg limit near the Mott boundary, as discussed 
previously~\cite{Parcollet99,GPS}.
This discussion thus certainly vindicates our chosen theoretical framework for
describing spin-liquid effects.
Finally, we also note that most recent studies on this general topic have focused on 
the more challenging finite-dimension Mott transition towards a spinon
liquid~\cite{Florens04,Lee05,Zhao07,Senthil08,Podolsky09,Ko11,Wang11,Potter12}.
Our study aims rather at including explicitly the effect of random magnetic exchange terms, 
which are notoriously difficult to address on finite dimensional lattice.

In order to investigate the interplay of Coulomb repulsion $U$ and exchange coupling 
$J$ close to the Mott metal-insulator transition, one needs to analyze the relative 
stability of the various phases.
This can be achieved from the knowledge of the total free energy (per site) which 
decomposes into two parts $F_{tot}=F_{imp}+F_{lat}$, respectively related to the local 
impurity problem (\ref{action}) solved by the rotor equations~(\ref{SX}-\ref{Sf}) and to
the original lattice model~(\ref{model}) via the self-consistency
equations~(\ref{selfconsistency1}-\ref{selfconsistency2}). The former
contribution can be obtained straightforwardly following the derivation of
the slave-rotor equation as saddle-point equations \cite{Florens02}:
\begin{eqnarray}
\label{Fimp}
F_{imp} &=& -\Lambda + \frac{N}{\beta}\sum_{\omega_n}[\ln\gfi-i\omega_n\gfi]\\
&&+\frac{1}{\beta}\sum_{\omega_n}[\ln\gxi-(\omega_n^2/U+\Lambda)\gxi]\nonumber\\
&&-N \int_0^\beta\!\!\! d\tau [\Delta(\tau)\gx\gf +\frac{1}{2} Q(\tau)\gf^2].\nonumber
\end{eqnarray}
Taken as a functional of the pseudoparticle Green's functions, the previous
expression generates the expected $\Sigma_X$ and $\Sigma_f$ self-energies 
Eqs.~(\ref{SX}-\ref{Sf}) by differentiation with respect to $G_X$ and $G_f$ 
respectively. The lattice contribution, associated to the DMFT equations,
can similarly be obtained from free energy functionals \cite{Georges04}:
\begin{eqnarray}
F_{lat} = \frac{N}{2}\int d\tau \frac{\Delta(\T)\Delta(\T)}{t^2}
+\frac{N}{4}\int d\tau \frac{Q(\T) Q(\T)}{J^2}.
\label{Flatt}
\end{eqnarray}
Minimization of $F_{tot}$ with respect to the dynamical fields $\Delta(\tau)$ 
and $Q(\tau)$ reproduces indeed the DMFT self-consistency condition 
Eqs.~(\ref{selfconsistency1}-\ref{selfconsistency2}).

\section{Discussion of the results}
\begin{figure}[ht]
 \includegraphics[width=8.5cm]{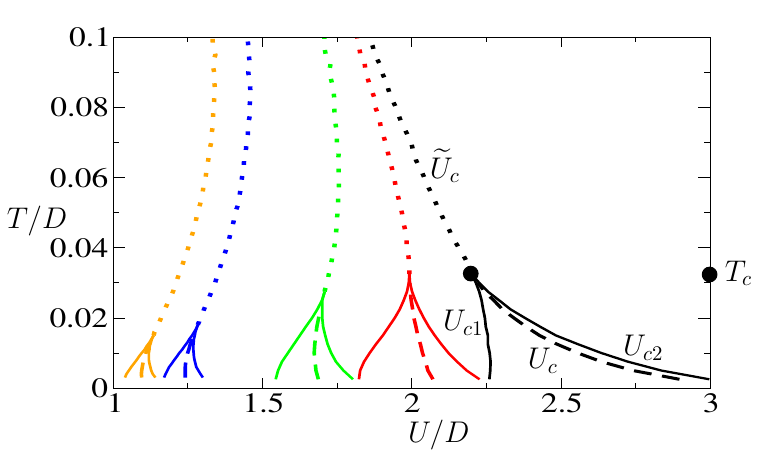}
\caption{Metal-insulator transition phase diagram for several values of
$J=0,0.1,0.2,0.4,0.5$ (right to left). Labels of the various critical lines, as defined 
in the text, are explicited for $J=0$. The absence of data for $T/D<0.003$ is
due to the simulations performed at finite temperature using Matsubara Green's
functions, but the explored temperature range covers all the interesting physical
regimes.}
\label{phase}
\end{figure}

We are now equipped to investigate how magnetic correlations, controlled
by the exchange term $J$, affect the Mott metal to insulator transition. 
To do so, we first construct the phase diagram in the ($U$,$T$) 
plane for fixed values of the magnetic exchange interaction $J$. 
At low temperatures this is done by monitoring the hysteresis in the local 
$d$-electron Green's function ($G_d(\tau)$ shows either slow Fermi liquid temporal 
decay or fast gapped behavior in the metallic and insulating states
respectively). This gives rise to two boundary lines $U_{c1}(T)$ and $U_{c2}(T)$, 
such that a single metallic (resp. insulating) solution exists at a given temperature 
$T$ for $U<U_{c1}(T)$ (resp. $U>U_{c2}(T)$). Furthermore, we determine from 
Eqs.~(\ref{Fimp}-\ref{Flatt}) the critical line $U_c(T)$ at which the free energies 
of metallic and insulating branches coincide.
The resulting phase diagram obtained for several values of $J$ from $J=0$ to $J/D=0.5$ is
displayed on Fig.~\ref{phase} in the case of the spin-orbital index $N=3$. We
stress that the particular choice of $N=3$ allows to match precisely the
phase boundary of the usual single-band Hubbard model for canonical spin-1/2
fermions~\cite{Florens02}, and is helpful to make quantitative comparisons when 
introducing other degrees of freedom (more orbitals for example) or new energy scales 
(random spin exchange as in the present work). We see from Fig.~\ref{phase} that 
irrespective of $J$, there always exists an interaction range $U_{c1}(T)<U<U_{c2}(T)$
bounding a coexistence region. With increasing temperature, the transition lines
always merge onto a critical point that occurs at  $T_c\simeq D/30$ for $J=0$, consistent with 
Refs.~\cite{Georges96,Florens02}.
Interestingly, Fig.~\ref{phase} shows that turning on the Heisenberg interaction
$J$ depresses both $T_c$ and $U_c$, leading also to a shrinking of the co-existence region. 
This can be understood by an energetic stabilization of the Mott phase via magnetic exchange. 
Turning to the zero temperature limit, one recovers by the collapse of the
$U_c(T)$ and $U_{c2}(T)$ lines at $J=0$ the continuous metal-insulator
transition described by the Brinkman-Rice scenario \cite{Georges96,Florens04}. 
In that case, the metal is the true ground state for all $U<U_{c2}(T=0)$. 
However, as seen in Fig.~\ref{phase}, the switching of $J$ markedly separates the actual critical
line $U_c(T)$ from the two stability lines, so that the Mott transition becomes first order at 
zero temperature, in agreement with recent findings on models with short-range
magnetic correlations via cluster-DMFT~\cite{Balzer08}.

Next, we turn our attention to the orientation of the transition lines as the interaction 
between the local-moments is increased.
At $J=0$, the $U_c(T)$ line is such that heating the correlated metal always leads
to the Mott insulator, a Pomeranchuk-type effect that relates to the high
entropy of the insulating phase.
This feature persists even {\it above} the transition temperature:
by determining at fixed $T>T_c$ the crossover interaction $\widetilde{U}_c(T)$ at
which the $d$-electron Green's function (taken e.g. at $\tau=\beta/2$) displays 
an inflection point, one distinguishes between a bad metal (i.e. an 
increasing resistivity with increasing temperature with non quadratic behavior) 
and a bad insulator (i.e. a non-gapped decreasing resistivity with increasing 
temperature), see \cite{Pruschke1993,Rozenberg1996,Merino2000,Georges03}. The continuity 
of the critical $U_c(T)$ line
(determined from the free energy at $T<T_c$) into the crossover line
$\widetilde{U}_c(T)$ (determined from the Green's function at $T>T_c$) serves as a 
vindication of our methodology. We also note that such a crossover at $T>T_c$ is
seen experimentally in the magnetic and transport measurement of $\kappa$-BEDT salts
\cite{Lefebvre00,Limelette03}.
Now, as $J$ is increased one expects the entropy of the paramagnetic Mott state to be 
reduced by magnetic exchange, thus weakening the Pomeranchuk lineshape. This is
indeed seen in Fig.~\ref{phase} by a progressive leftward bending of the critical lines 
at increasing $J$, and consistent with results of
Ref.~\cite{Park08,Balzer08,Liebsch09}.
A particularity of our model is the re-entrance of the Pomeranchuk effect at the
lowest observable temperature, which we could attribute to the reduced (with respect to 
$\log(2)$ of free moments) yet non-zero entropy of the insulating spin liquid
state \cite{Parcollet99,Sachdev93}. We have checked also that increasing orbital
degeneracy with larger $N$ values tends to weaken all magnetic effects,
recovering the standard $J=0$ phase diagram.

\begin{figure}[ht]
\includegraphics[width=8.7cm]{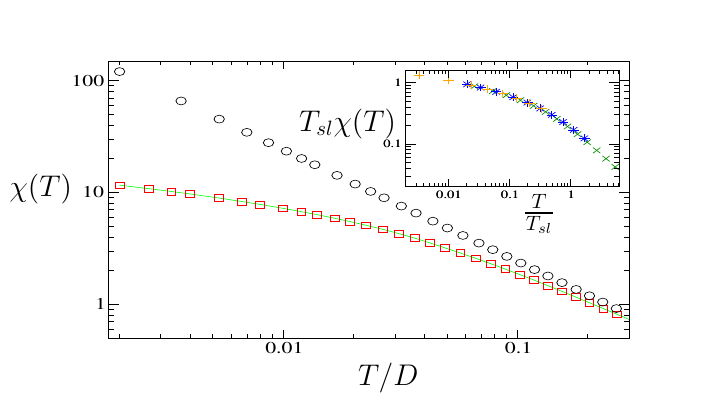}
\caption{Spin susceptibility $\chi(T)$ versus temperature for $J=0$ and large
$U$ (circles, corresponding to uncoupled magnetic moments in the Mott phase), and for 
$J=0.2$ and $U/D=1.8$ (squares, corresponding to a spin-liquid Mott phase). 
A $1/T$ Curie behavior applies to the whole $J=0$ curve (free moments), while a spin-liquid 
susceptibility Ansatz $\chi(T)=(1/4T_{sl})\ln(1+T_{sl}/T)$
describes the case of finite $J=0.2$ exchange (full line). The inset
demonstrates scaling of our numerical data for three different values of the exchange
interaction $J=0.2,0.4,0.8$ in the spin-liquid Mott phase.}
\label{chi}
\end{figure}

\begin{figure}[ht]
\includegraphics[width=8.5cm]{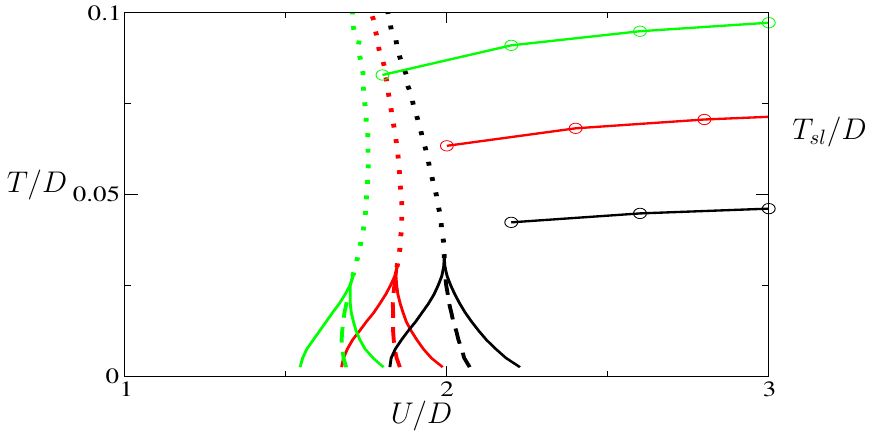}
\caption{Superposed plot of the $U$-dependent spin-liquid onset temperature $T_{sl}$ 
for several $J=0.1,0.15,0.2$ (circles, bottom to top) with the corresponding metal-insulator 
phase diagrams (right to left). The merging of $T_{sl}$ with the temperature
scale where the crossover lines bend demonstrate the crucial role of the
magnetic fluctuations in the destruction of the Pomeranchuk effect.}
\label{Tsl}
\end{figure}

A key point of our study concerns the connection between the exchange induced
modifications of the phase diagram (which can be observed e.g. in
transport experiments) and the anomalous spin-liquid dynamics.
The spin susceptibility $\chi(\T)$ in the absence of exchange $J=0$ reduces
trivially to a pure Curie law in the whole Mott state $U>U_{c2}$, as checked in
Fig.~\ref{chi}.
At finite $J$, Curie behavior is seen only at high temperatures such that $T\gg
J$, while departure from the independent local moment picture is observed by a
slower logarithmic divergence at zero temperature.
A crossover scale $T_{sl}$ characterizing the onset of the spin liquid state
can be accurately determined from a fit of $\chi(T)$ by the appealing
form $ \chi(T)=1/(4T_{sl})\ln(1+T_{sl}/T)$ which follows the expected high and low
temperature regimes (the agreement of this Ansatz with our data is excellent,
see inset of Fig.~\ref{chi} where scaling is demonstrated).
This simple Ansatz allows us to extract precisely, for a given value of $J$, the $U$-dependence 
of the spin-spin correlation scale $T_{sl}$, see Fig.~\ref{Tsl}, obtained for 
several $J$ values. 
We note first that $T_{sl}$ saturates at very large $U$ values to (half) the bare magnetic exchange
constant $J$, an expected result since only spin dynamics plays a role upon
freezing of the charge excitations, leaving $J$ as the only relevant scale.
More interesting is the net decrease of $T_{sl}$ when $U$
is diminished towards the critical Mott boundary. This feature signals a weakening
of the spin correlations in the vicinity of the Mott transition, brought by
the finite probability of double and zero occupation of the $d$ state
(valence fluctuations in the particle-hole symmetric regime).
Quite intriguing is also that the spin susceptibility continues to display the
characteristic signature of the spin liquid even in a regime where the
ground state is metallic, close to the coexistence region, as seen by the
fact that $T_{sl}$ remains finite near the Mott boundary (note that it is
difficult to determine $T_{sl}$ precisely in the metallic phase, due to the
existence of multiple energy scales). This result is reminiscent of a previous study 
by Parcollet and Georges \cite{Parcollet99}, where 
the doping-induced Mott transition (at $U=\infty$) within the fully random Heisenberg 
model was considered. These authors showed that the magnetic crossover scale $T_{\rm sl}$ 
persists at small doping, leading to a slush state of incoherent metal
with spin liquid correlations, similar to what we observe here at half-filling
in the intermediate Coulomb interaction regime.
%
Our last and perhaps most important prediction is the existence of a clear link 
between the spin liquid coherence scale $T_{sl}$ and the temperature where the bending
of the metal-insulator crossover lines takes place, as manifest from Fig.~\ref{Tsl}.
This crucial observation indeed shows that magnetic exchange is responsible for the quenching 
of the Pomeranchuk effect and the bending of the Mott transition lines, a result
previously anticipated on physical grounds.

\section{Conclusion}
We have investigated the role played by fluctuating local moments coupled 
via frustrated long-range interactions near the Mott boundary.
The metal-insulator transition was found to be modified by the Heisenberg exchange in favor 
of the insulating state and even to turn discontinuous at zero temperature.
Studying the magnetic susceptibility, we were able to connect the bending of the critical lines 
to the appearance of collective spin correlations, with reduced effective
exchange. These effects should be generic to other microscopic models and experimental 
systems near the Mott transition. One interesting open question is whether
the quantum critical behavior of the large dimension Mott transition~\cite{Terletska11}
persists to a quenching of magnetic moments induced by exchange interactions.
Recent development of efficient bosonic impurity solvers~\cite{Bulla,Freyn,Guo}
should allow such further explorations of this rich physics.

We would like to acknowledge helpful discussions with V. DOBROSAVLJEVIC, S. R. HASSAN, 
R. MOESSNER, and T. VOJTA.


\begin{thebibliography}{99}

\bibitem{Imada98} IMADA M., FUJIMORI A. and TOKURA Y.,
{\it Rev. Mod. Phys.}, {\bf 70} (1998) 1039.
\bibitem{Georges96} GEORGES A., KOTLIAR G., KRAUTH W. and ROZENBERG M. J.,
{\it Rev. Mod. Phys.}, {\bf 68} (1996) 13.

\bibitem{Lefebvre00} LEFEBVRE S., WZIETEK P., BROWN S., BOURBONNAIS C.,
J\'EROME D., M\'EZI\`ERE C., FOURMIGU\'E M. and BATAIL P., {\it Phys. Rev.
Lett.}, {\bf 85} (2000) 5420.
\bibitem{Limelette03} LIMELETTE P., WZIETEK P., FLORENS S., GEORGES A., 
COSTI T. A., PASQUIER C., J\'EROME D., M\'EZI\`ERE C. and BATAIL P., {\it Phys. 
Rev.  Lett.} {\bf 91} (2003) 016401.
\bibitem{Kagawa04} KAGAWA F., ITOU T., MIYAGAWA K. and KANODA K., {\it Phys. Rev.
B} {\bf 69} (2004) 064511.

\bibitem{Ohashi08} OHASHI T., MOMOI T., TSUNETSUGU H. and KAWAKAMI N., {\it Phys.
Rev. Lett.} {\bf 100} (2008) 076402.
\bibitem{Park08} PARK H., HAULE K. and KOTLIAR G., {\it Phys. Rev. Lett.} {\bf
101} (2008) 186403.
\bibitem{Balzer08} BALZER M., KYUNG B., S\'EN\'ECHAL D., TREMBLAY A.-M. S. and 
POTTHOFF M., {\it Europhys. Lett.} {\bf 85} (2009) 17002.
\bibitem{Liebsch09} LIEBSCH A., ISHIDA H. and MERINO J., {\it Phys. Rev. B} {\bf
79} (2009) 195108.

\bibitem{Vlad} AGUIAR M. C. O. and DOBROSAVLJEVIC V., {\it Phys. Rev. Lett.} 
{\bf 110} (2013) 066401.

\bibitem{Sachdev93} SACHDEV S. and YE J., {\it Phys. Rev. Lett.} {\bf 70} (1993) 3339.

\bibitem{Parcollet99} PARCOLLET O. and GEORGES A., {\it Phys. Rev. B} {\bf 59}
(1999) 5341.

\bibitem{RSidd} GEORGES A., SIDDHARTHAN R. and FLORENS S.,
{\it Phys. Rev. Lett.} {\bf 87}, (2001) 277203.

\bibitem{Rosch} HAULE K., ROSCH A., KROHA J. and WOELFLE P.,
{\it Phys. Rev. Lett.} {\bf 89}, (2002) 236402.

\bibitem{Miranda} MIRANDA E. and DOBROSAVLJEVIC V., {\it Rep. Prog. Phys.} {\bf 68} 
(2005) 2337.

\bibitem{Hassan13} HASSAN S. R. and S\'EN\'ECHAL D., {\it Phys. Rev. Lett.} {\bf 110}
(2013) 096402.

\bibitem{Bray80} BRAY A. J. and MOORE M. A., {\it J. Phys. C: Solid State Phys.}, {\bf
13} (1980) L655.

\bibitem{Auerbach94} AUERBACH A. {\it ``Interacting electrons and quantum
magnetism''}, Springer (1994).

\bibitem{Florens02} FLORENS S. and GEORGES A., {\it Phys. Rev. B} {\bf 66}
(2002) 165111.
\bibitem{Florens04} FLORENS S. and GEORGES A., {\it Phys. Rev. B} {\bf 70}
(2004) 035114.

\bibitem{Grempel} GREMPEL D. R. and ROZENBERG M. J., {\it Phys. Rev. Lett.} 
{\bf 80} (1998) 389.
\bibitem{Arrachea} ARRACHEA L. and ROZENBERG M. J., {\it Phys. Rev. Lett.} 
{\bf 86} (2001) 5172.
\bibitem{GPS} GEORGES A., PARCOLLET O. and SACHDEV S., {\it Phys. Rev. Lett.} 
85 (2000) 840.
\bibitem{VBS} VOJTA M., BURAGOHAIN C. and SACHDEV S. {\it Phys. Rev. B} {\bf 61}
(2000) 15152.

\bibitem{Lee05} LEE S.-S. and LEE P. A., {\it Phys. Rev. Lett.} {\bf 95} 
(2005) 036403.
\bibitem{Zhao07} ZHAO E. and PARAMEKANTI A., {\it Phys. Rev. B} {\bf 76} 
(2007) 195101.
\bibitem{Senthil08} SENTHIL T., {\it Phys. Rev. B} {\bf 78} (2008) 045109.
\bibitem{Podolsky09} PODOLSKY D., PARAMEKANTI A., KIM Y. B. and SENTHIL T.,
{\it Phys. Rev. Lett.} {\bf 102} (2009) 186401.
\bibitem{Ko11} KO W.-H. and LEE P. A., {\it Phys. Rev. B} {\bf 83} 
(2011) 134515.
\bibitem{Wang11} WANG G., GOERBIG M. O., MINIATURA C. and GR\'EMAUD B.,
{\it Europhys. Lett.} {\bf 95} (2011) 47013.
\bibitem{Potter12} POTTER A. C., BARKESHLI M., McGREEVY J. and SENTHIL T.,
{\it Phys. Rev. Lett.} {\bf 109} (2012) 077205.

\bibitem{Georges04} GEORGES A., {\it ``Lectures on the Physics of Highly Correlated 
Electron Systems VIII''} (2004), American Institute of Physics Conference Proceedings 
Vol. {\bf 715}, also preprint arXiv:cond-mat/0403123.


\bibitem{Pruschke1993} PRUSCHKE T., COX D. L. and JARRELL M., {\it 
Phys. Rev. B} {\bf 47} (1993) 3553.
\bibitem{Rozenberg1996} ROZENBERG M. J., KOTLIAR G. and KAJUETER H., 
{\it Phys. Rev. B} {\bf 54}, (1996) 8452.
\bibitem{Merino2000} MERINO J. and McKENZIE
R. H., {\it Phys. Rev. B} {\bf 61}, (2000) 7996.
\bibitem{Georges03} GEORGES A., FLORENS S. and COSTI T. A., {\it J. Phys.  IV} {\bf 114}
(2004) 165.
\bibitem{Terletska11} TERLETSKA H., VUCICEVIC J., TANASKOVIC D. and
DOBROSAVLJEVIC V., {\it Phys. Rev. Lett.} {\bf 107} (2011) 026401.ć

\bibitem{Bulla} BULLA R., LEE H. J., TONG N. H. and VOJTA M., {\it Phys. Rev. B} 
{\bf 71} (2005) 045122.
\bibitem{Freyn} FREYN A. and FLORENS S., {\it Phys. Rev. B} {\bf 79} (2009) 121102.
\bibitem{Guo} GUO C., WEICHSELBAUM A., von DELFT J. and VOJTA M., {\it
Phys. Rev. Lett.} {\bf 108} (2012) 160401.




\end{thebibliography}
\end{document}